# Rheological Signatures of Gelation and Effect of Shear Melting on Aging Colloidal Suspension

Shweta Jatav and Yogesh M Joshi*

Department of Chemical Engineering

Indian Institute of Technology Kanpur, INDIA

* Corresponding author, E-mail: joshi@iitk.ac.in

# Synopsis

Colloidal suspensions that are out of thermodynamic equilibrium undergo physical aging wherein their structure evolves to lower the free energy. In aqueous suspension of Laponite, physical aging accompanies increases of elastic and viscous moduli as a function of time. In this work we study temporal evolution of elastic and viscous moduli at different frequencies and observe that freshly prepared aqueous suspension of Laponite demonstrates identical rheological behavior reported for the crosslinking polymeric materials undergoing chemical gelation. Consequently at a certain time tan$\delta$ is observed to be independent of frequency. However, for samples preserved under rest condition for longer duration before applying the shear melting, the liquid to solid transition subsequent to shear melting shows greater deviation from classical gelation. We also obtain continuous relaxation time spectra from the frequency dependence of viscous modulus. We observe that, with increase in the rest time, continuous relaxation time spectrum shows gradual variation from negative slope, describing dominance of fast relaxation modes to positive slope representing dominance of slow relaxation modes. We propose that the deviation from gelation behavior for the shear melted suspensions originates from inability of shear melting to completely break the percolated structure thereby creating unbroken aggregates. The volume fraction of such unbroken aggregates increases with the rest time. For small rest times presence of fewer number of unbroken aggregates cause deviation from the classical gelation. On the other hand, at high rest times presence of greater fraction of unbroken aggregates subsequent to shear melting demonstrate dynamic arrest leading to inversion of relaxation time spectra.

# I. Introduction

In colloidal suspensions inter-particle interactions and the shape of particles play an important role in determining ultimate microstructure of the same. For predominantly repulsive interactions, beyond a certain concentration, suspended particles can get arrested in a disordered state if crystallization is avoided [Hunter and Weeks (2012)]. Such disordered state is also known as colloidal glass, wherein density is homogeneous beyond a length-scale of the order of inter-particle distance [Tanaka *et al.* (2004)]. On the other hand, if inter-particle interactions are primarily attractive, suspended particles form aggregates having a fractal network like structure. When such aggregates are sufficiently large to touch each other and span the space, colloidal gelation occurs [Coussot (2007), Joshi (2014)]. In a colloidal gel, density is homogeneous beyond an aggregate length-scale, which is much larger than the particle length-scale [Tanaka et al. (2004)]. Although, in principle, the microstructure of a glass state in comparison with a gel state is distinctly different, materials in both these states show many similar physical and rheological behaviors, and have been collectively termed as soft glassy materials [Fielding *et al.* (2000)]. Usually this class of materials are thermodynamically out of equilibrium as constrained mobility of the particles arrested in these states does not allow complete access to their phase space [Liu and Nagel (1998)]. Consequently, owing to thermal motion of the particles, their microstructure evolves to lower the free energy as a function of time, a process typically known as physical aging [Joshi (2014), Viasnoff *et al.* (2003)]. Application of deformation field, on the other hand, destroys the structure formed during the physical aging enhancing the free energy of the material, a process normally known as rejuvenation or shear melting [Joshi *et al.* (2008), McKenna *et al.* (2009), Viasnoff et al. (2003)]. Structural reorganization or physical aging resumes upon cessation of shear melting, a property known as thixotropy [Mewis and Wagner (2009)]. The deformation field, therefore, has a profound effect on the structural reorganization in this class of materials. Usually a gel state and a glassy state are characterized using microscopic and scattering techniques, though the rheological techniques have also been employed recently to distinguish the same [Habdas and Weeks (2002), Krishna Reddy *et al.* (2012), Laurati *et al.* (2011), Mohraz and Solomon (2006), Negi *et al.* (2014), Pham *et al.* (2008), Ruzicka *et al.* (2008), Shahin and Joshi (2012), Shao *et al.* (2013), Verduin *et al.* (1996), Winter (2013)]. In this work, we study effect of shear melting on colloidal gelation and the subsequent dynamics in aging aqueous suspension of Laponite clay.

The dominating attractive interactions among the colloidal particles lead to the formation of fractal aggregates that form the building blocks of a colloidal gel. If attractive potential is very strong, aggregates are more open with fractal dimension 1.7 to 1.9 (diffusion limited cluster aggregation). On the other hand, if multiple collisions are necessary before forming a bond, more compact aggregates are formed with fractal dimension in the range 2 to 2.1 (reaction limited cluster aggregation) [Lu and Weitz (2013)]. Interestingly dynamic light scattering of colloidal gel shows that the intermediate scattering function does not decay to zero after gelation transition [Pusey *et al.* (1993)]. In various kinds of colloidal particles, the inter-particle interaction can be tuned to show the temperature dependence. Such particles demonstrate thermo-reversible gelation [Larson (1999)]. Gelation of anisotropic particles having different aspect ratios and interaction potentials has also been studied in the literature. For prolate particles this subject has been recently reviewed by Solomon and Spicer (2010). The work on oblate particles essentially comprises of various kinds of clay suspensions, wherein strength of charges and their distribution on the particle in addition to the aspect ratio influences the phase behavior [Luckham and Rossi (1999), Ruzicka and Zaccarelli (2011), Van Olphen (1977)].

The significant work on the colloidal glasses has been carried out on spherically shaped particles that interact via hard sphere interactions, wherein potential is zero if the particles are not touching but otherwise infinity [Hunter and Weeks (2012)]. In such system, since the potential energy associated with any configuration is zero, maximization of entropy characterizes the equilibrium state. On the other hand, many systems comprising of hairy particles and multiarm star polymer suspension/melt also demonstrate behavior similar to hard sphere colloidal glasses with extra relaxation modes originating from arm/hair dynamics [Agarwal *et al.* (2009), Rogers *et al.* (2010)]. Typically below the volume fraction of 0.494 suspension of monodispersed spherical particles is in liquid state, while that of 0.545 colloidal crystalline state is formed [Hunter and Weeks (2012)]. If crystallization is prevented by rapidly increasing concentration or by inducing slight polydispersity, suspension lies in colloidal glassy region over a volume fraction range: 0.56 to 0.64 [Hunter and Weeks (2012), Larson (1999)]. van Megen and Pusey (1991) were the first to verify this who carried out dynamic light scattering experiments on sterically stabilized PMMA suspension which closely follows hard sphere interactions. They observed that intermediate scattering function decay to zero for volume fractions only below 0.56 indicating ergodicity breaking and glassy behavior for higher concentration suspensions. Ergodicity breaking in colloidal glasses is attributed to caging effect wherein particles are arrested to cage like environment by the surrounding particles, thereby preventing them from accessing the phase space [Joshi (2014)]. The direct visualization of the structural rearrangement in colloidal glasses has been facilitated by various microscopic techniques [Hunter and Weeks (2012), Lu and Weitz (2013)]. Similar to spherical particles, suspension of anisotropic particles also leads to formation of glasses. However, the critical volume fraction beyond which glassy state is formed decreases with increase in extent of anisotropy [Solomon and Spicer (2010)]. The physical cages are also present in microgel pastes [Cloitre *et al.* (2000), Di *et al.* (2011)], concentrated emulsions [Mason *et al.* (1995)] and foams [Cipelletti and Ramos (2005)]. Although, unlike particulate glasses, dispersed phase in these systems can get deformed leading to volume fractions approaching unity, many dynamic behaviors of the same are reminiscent of particulate colloidal glasses.

Rheologically both the colloidal gels and glasses demonstrate viscoelasticity, however origin of elasticity in glasses is due to caging effect while that of in gels is due to strength of backbone fractal network in aggregates [Coussot (2007)]. In addition to aging and rejuvenation, colloidal glasses and gels are observed to demonstrate more complex phenomena such as overaging [Bandyopadhyay *et al.* (2010), Viasnoff et al. (2003)] and shear banding [Besseling *et al.* (2010), Paredes *et al.* (2011)].

In this work we study aqueous suspension of Laponite, which is synthetic clay belonging to a family of hectorite. Primary particle of Laponite is disk shaped with diameter 30 nm and thickness 1 nm [Kroon *et al.* (1996)]. In an aqueous media, faces of Laponite particle have permanent negative charge [Ruzicka and Zaccarelli (2011)]. To avoid dissolution of Laponite in water, pH is usually maintained in the range 9 to 10 [Thompson and Butterworth (1992)]. Over this pH range charge on the edges of Laponite particles is positive [Tawari *et al.* (2001)]. Consequently Laponite particles in aqueous media experiences face - to - face repulsive interactions, while edge - to - face attractive interactions [Cummins (2007), Mongondry *et al.* (2005), Ruzicka and Zaccarelli (2011)]. Addition of monovalent salt such as NaCl in suspension enhances ionic concentration, which shields the charges on the Laponite particles thereby reducing the repulsive interactions [Shahin and Joshi (2012)]. Owing to anisotropic shape and dissimilar charge distribution, aqueous suspension of Laponite shows a very rich phase behavior that evolves as a function of time [Cummins (2007), Mongondry et al. (2005), Mourchid *et al.* (1995), Ruzicka and Zaccarelli (2011), Shahin *et al.* (2011)]. There is a consensus in the literature that aqueous suspension of Laponite, below around 2 weight %, forms an attractive gel having a house of cards structure formed by edge - to - face interactions [Ruzicka and Zaccarelli (2011)]. The microstructure of Laponite suspension above concentration of 2 weight % is, however, a matter of debate [Cummins (2007), Mongondry et al. (2005)]. The two proposals exist in the literature. According to the first proposal attractive gel observed below 2 weight % concentration continues to dominate at higher concentrations as well [Cocard *et al.* (2000), Mongondry et al. (2005)]. On the other hand, the second proposal considers repulsion between the faces of Laponite particles giving rise to repulsive or Wigner glass like microstructure [Jabbari-Farouji *et al.* (2008), Ruzicka and Zaccarelli (2011)].

Irrespective of whether a gel or a glass, aqueous suspension of Laponite shows prominent thixotropic behavior; wherein under quiescent conditions it undergoes physical aging; while under application deformation field it undergoes rejuvenation [Bonn *et al.* (2004), Joshi et al. (2008), Willenbacher (1996)]. Typically during the physical aging, the mean relaxation time and the elastic modulus of Laponite suspension show a prominent increase as a function of time. On the other hand, during the rejuvenation the same variables decrease [Negi and Osuji (2009), Shahin and Joshi (2012)]. Furthermore, owing to anisotropic shape, complex inter-particle interactions and constrained mobility at the particle level, application of deformation field explores those sections of energy landscape (or forms such structures), which otherwise are not accessible without the deformation field [Bandyopadhyay et al. (2010), Joshi *et al.* (2012)]. Typically flow induced nematic alignment of Laponite particles is observed due to shear melting [Shahin and Joshi (2010)]. In addition, Shahin and Joshi (2010) observed that shear melting does not rejuvenate an older gelled samples (kept under quiescent conditions for a longer duration) of Laponite suspension to the state same as that of freshly prepared suspension. This suggests physical aging in Laponite suspension is partly irreversible, wherein strong deformation field applied during shear melting cannot completely obliterate the structure evolved during aging. Owing to all these reasons, structural reorganization during physical aging followed by shear melting is different from structural organization in freshly prepared suspensions. Recently Angelini *et al.* (2013) employed dynamic light scattering to investigate relaxation dynamics of Laponite suspensions with and without shear melting. Interestingly they observed that autocorrelation function of the spontaneously aged suspension shows stretched exponential decay, while the shear melted suspension shows compressed exponential decay.

Comparable to liquid – solid transition observed in colloidal suspensions, polymeric liquids are known to undergo transformation from liquid to apparent solid-like state either by glass transition (rapidly decreasing temperature as a function of time) or by chemical gelation (crosslinking reaction). Rheologically both these processes demonstrate very distinct signatures in small amplitude oscillatory experiments. Particularly a glass transition point is characterized by peak in $\tan\delta$, which is a ratio of viscous modulus ($G''$) to elastic modulus ($G'$), as temperature is decreased [Shaw and MacKnight (2005)]. However, the glass transition point depends on observation timescale (inverse of frequency), and the corresponding glass transition temperature $(T_g)$ is known to increase with increase in frequency [Becker (1955)]. On the other hand, in a crosslinking reaction, the chemical gel point (GP), defined as a critical state at which mass average molecular weight diverges, and is obviously independent of timescale of observation. Winter and Chambon (WC) suggested that at GP, $G'$ and $G''$ demonstrate following dependence on frequency ($\omega$) [Winter and Chambon (1986)]:

$$G' = G'' \cot(n\pi/2) = \frac{\pi S}{2\Gamma(n)\sin(n\pi/2)} \omega^n, \qquad (1)$$

where $n$ is the relaxation exponent $(0 < n < 1)$ and $S$ is the gel strength. Typically in the crosslinking polymeric materials approaching GP, $\tan \delta$ is observed to respectively decrease and increase with frequency indicating a sol and a gel state. However at GP, the loss angle $\delta = n\pi/2$, is observed to be independent of frequency. Winter (1987) suggested that for a stoichiometrically balanced end-linking network $n=0.5$, while for excess (stiff gel) and dearth (soft gel) of crosslinker, $n$ respectively decreases or increases. Furthermore, there are several reports available that lead to estimation of fractal dimension of an aggregate from $n$ [Cocard et al. (2000), Muthukumar (1989), Ponton et al. (2005)]. Recently Winter (2013) proposed that glass transition is the rheological inverse of gelation. He compared continuous relaxation time spectra at GP in a crosslinking polymeric system with that of a concentrated colloidal suspension approaching the glassy state. He observed that the spectrum has a negative slope at GP, while a positive slope at the glass transition. He argued that such behavior originates from dominance of small time-scales at GP unlike that of material approaching glass.

The WC criterion that at chemical GP $\tan \delta$ does not depend on frequency has also been observed for physical gels, where, in principle, the network is formed by reversible interactions. However, it is not necessary that all the suspensions that show gelation transition should show GP [Krishna Reddy et al. (2012)]. There are various studies that observe and analyze GP for a number of thermo-reversible gels such as: syndiotactic Poly vinyl alcohol solution [Choi *et al.* (2001)], gelatin [Hodgson and Amis (1991), Hsu and Jamieson (1993)], aqueous suspension of rod-like particles with grafted poly(N-isopropylacrylamide) [Krishna Reddy et al. (2012)], colloidal silica with octadecyl chains in decalin [Negi et al. (2014)], etc. These studies report gelation transition to demonstrate the critical state (GP), wherein $G'$ and $G''$ are observed to follow equation (1). Compared to thermo-reversible gels, there is lesser number of reports that show GP for other types of physical gels. Ponton et al. (2005) , studied physical gelation of charged magnetic nanoparticles in aqueous media and observed $n$ to vary between 0.25 to 0.6 for different concentrations and pH conditions. Cocard and coworkers (2000) studied 1 weight % aqueous suspension of Laponite and observed gelation to be consistent with WC criterion with $n=0.55$. They propose that value of the exponent suggests percolated network with elastic response dominated by bending modes and screened hydrodynamic interactions. However, in an apparent contradiction to the observation by Cocard and coworkers (2000) for 1 weight % Laponite suspension, Negi and coworkers (2014) did not observe any GP for 4 days old 3.5 weight % shear melted Laponite suspension. They conclude that the behavior is suggestive of colloidal glass undergoing dynamic arrest.

The observation of Cocard and coworkers (2000) in comparison with Negi and coworkers (2014) raise two important issues as follows. It is imperative to understand whether the rheological aging behavior of low (< 2 weight %) and high concentration Laponite suspension is different such that the former shows signature of GP [Cocard et al. (2000)] while the latter shows colloidal glass transition [Negi et al. (2014)], as suggested by the recently published phase diagram [Ruzicka and Zaccarelli (2011)]. In addition Laponite suspension is known to undergo irreversible aging as mentioned before [Shahin and Joshi (2010)]. Therefore do irreversible aging, in addition to shear melting, has any role to play in rheological signatures of structural evolution for higher concentration (> 2 weight %) Laponite suspensions. To address these questions, we analyze spontaneously aged as well as shear melted 2.8 weight % aqueous suspension of Laponite having 3 mM of NaCl by subjecting the same to cyclic frequency sweep.

This system has been proposed to be forming a repulsive Wigner glass [Ruzicka and Zaccarelli (2011)]. We investigate behavior of $G'$, $G''$ and $\tan\delta$ as a function of frequency and study the evolution of continuous relaxation time spectrum as the suspension undergoes physical aging.

## II. Sample preparation and experimental protocol

In this work we use 2.8 weight % Laponite XLG® suspension having 3 mM NaCl. Suspension is prepared by gradually adding oven dried (4 hour at 120°C) Laponite XLG in Millipore® water maintained at pH 10 by NaOH and containing 3 mM NaCl. The mixing was carried out using ultra Turex drive for 30 min before storing the same in sealed polypropylene bottles. We perform the experiments on samples kept under quiescent (rest) conditions for $t_R$ =0, 11, 22, 30, 144 h after preparation of the sample. This time is represented as a rest time $(t_R)$. Before starting each experiment, stored sample (except for 0 h, which is used immediately after preparation) is introduced in the shear cell using a dropper, and subjected to shear melting under oscillatory stress of 40 Pa and frequency of 0.2 Hz for 10 minutes. After cessation of shear melting, cyclic frequency sweep with stress magnitude of 0.1 Pa is applied to the samples over a frequency range of 0.063 – 20 rad/s. We represent the time elapsed since cessation of shear melting by aging time or waiting time $(t_w)$. In this work we use DHR 3 (TA Instruments) rheometer with a concentric cylinder geometry (cup diameter 30 mm and gap 1 mm). In order to prevent evaporation of water over the duration of the experiments a thin layer of silicon oil is applied on the free surface of the samples. All the experiments are carried out at 20°C.

In aqueous suspension of Laponite the rheological properties are time dependent. Consequently, representation of instantaneous rheological behavior $G'$ and $G''$ is reliable as long as properties do not evolve significantly over a period of single cycle under application of oscillatory flow field. Mours and Winter (1994) suggested that estimation of $G'$ and $G''$ is acceptable when mutation numbers given by:

$$N'_{mu} = \frac{2\pi}{\omega G'}\frac{\partial G'}{\partial t} \text{ and } N''_{mu} = \frac{2\pi}{\omega G''}\frac{\partial G''}{\partial t}, \qquad (2)$$

are small, typically $N'_{mu} < 0.1$ and $N''_{mu} < N'_{mu}$. We estimate the mutation numbers as a function of time for the data obtained for samples at various $t_R$, and observe that that except for few points in the limit of small frequency, the mutation numbers are well within the recommended limit.

### III. Results and Discussion

As discussed in the experimental procedure, every sample is subjected to shear melting before applying cyclic frequency sweep. In figure 1(a) we plot change in complex viscosity as a function of time under application of high magnitude of oscillatory stress in the shear melting step. It can be seen that immediately after preparation of the suspension, complex viscosity remains constant throughout the shear melting experiment. However the samples stored under quiescent condition after preparation, undergo physical aging, wherein microstructure develops as a function of time. Shear melting on such samples causes complex viscosity to decreases as a function of rejuvenation time, which eventually reaches a plateau. In the same figure, we also plot magnitude of oscillatory strain as a function of rejuvenation time. It can be seen that strain shows a dependence opposite of that of complex viscosity as a function of rest time $(t_R)$ and eventually attains a constant value. In figure 1(b), we plot the viscosity and strain associated with the plateau, which can be seen to be respectively increasing and decreasing with the rest time. This suggests that shear melting is not able to completely break the structure formed during the physical aging in aqueous suspension of Laponite. It is important to note that application of higher magnitude of stress does not rejuvenate the samples any further. This is because the aging subsequent to shear melting is observed to be unaffected by magnitude of stress applied during the shear melting, if it is sufficiently greater than the yield stress [Shahin and Joshi (2010), Viasnoff et al. (2003)]. In order to confirm this, shear melting is carried out on some samples with stress magnitude of 60 Pa. As expected, evolution of $G'$ and $G''$ subsequent to shear melting is observed to be independent of the magnitude of stress applied during shear melting within experimental uncertainty. Interestingly Shukla and Joshi (2009) and Joshi and coworkers (2012) reported that application of higher magnitude of stress, even though above the yield stress of the material at the time of application, cannot prevent slow reorganization of the structure over a very prolong timescale. In the present work we apply oscillatory shear stress over a period of 10 min, wherein we observe $\eta^*$ and $G''$ to demonstrate a plateau, while $G'$ is observed to be below the detection limit. Such state, therefore, for all the practical purposes, can be considered to be a completely shear rejuvenated state for a given structure of the material.

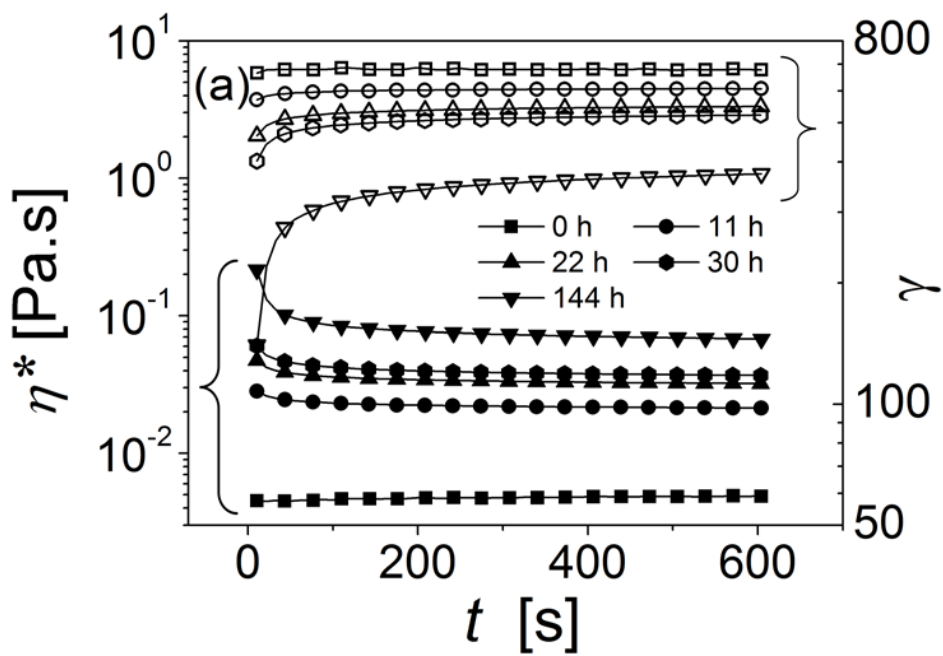

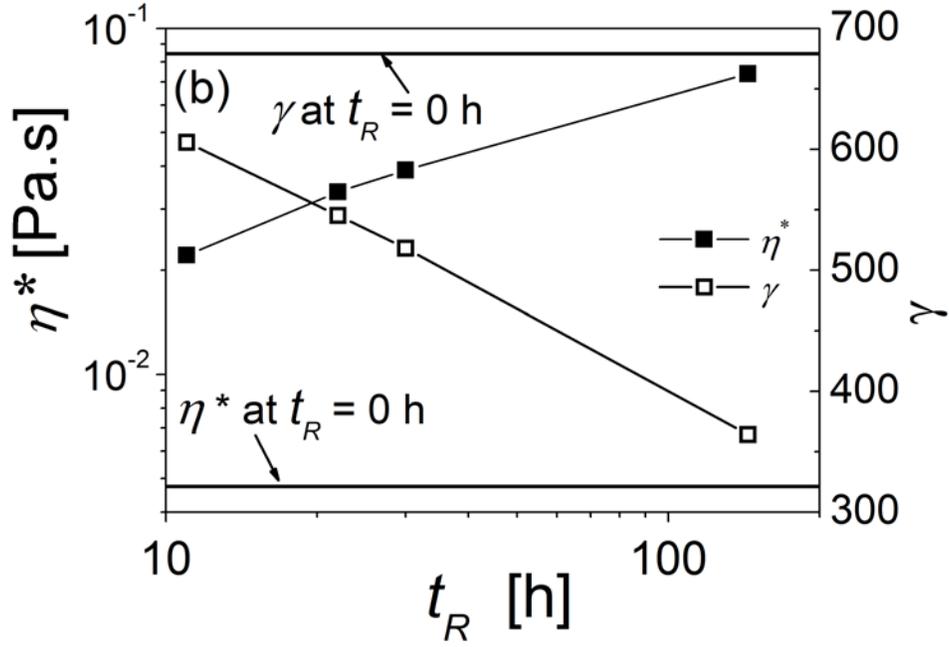

**Figure 1.** (a) Evolution of complex viscosity $\eta^*$ at left axis (filled symbols: from bottom to top for $t_R$ =0, 11, 22, 30 and 144 h) and variation of strain amplitude $\gamma$ during shear melting at right axis (open symbols; from top to bottom for $t_R$ =0, 11, 22, 30 and 144 h) (b) Plateau value of complex viscosity (left axis) and strain amplitude (right axis) shown in (a) plotted a as function of rest time . The lines connecting data are guide to the eye.

Subsequent to shear melting, the samples are subjected to cyclic frequency sweep, which enables estimation of $G'$ and $G''$ as a function of $t_w$ at the explored frequencies in a single experiment. In figure 2(a) we plot evolution of $G'$ and $G''$ with respect to $t_w$ at different frequencies for a freshly prepared ($t_R$ =0 h) sample. It can be seen that both the moduli increase with time with $G'$ increasing at faster rate than $G''$. At very small times the sample is in liquid state $(G' < G'')$, however eventually $G'$ crosses over $G''$. The time, at which crossover $(G' = G'')$ occurs, increases with decrease in $\omega$. We also plot the corresponding evolution of $\tan\delta$ in the same figure. It can be seen that $\tan\delta$ associated with a lower $\omega$ decreases at faster rate. However, at $\tan\delta \approx 0.68$ ($t_{GP}$=457 min), all the curves pass through a single point leading to $n \approx 0.38$ according to equation (1). Rheologically such point has been termed as the gel point (GP). The corresponding dependence of $G'$ and $G''$ on $\omega$ at different times elapsed since shear melting is plotted in figure 2(b). The frequency dependence plots at respective time intervals have been shifted horizontally for clarity; and the corresponding shift factors are mentioned in table 1. It can be seen that at small $t_w$, $G'$ and $G''$ demonstrates characteristic terminal behavior wherein dependence of $G' \propto \omega^2$ and $G'' \propto \omega$ is observed. We also plot $\tan\delta$ as a function of $\omega$ at different $t_w$, which at small $t_w$ shows prominent decrease with $\omega$, a characteristic feature associated with a sol state. However as time passes, the dependence of both the moduli and $\tan\delta$ on $\omega$ weakens, and at GP, both the moduli show identical dependence on $\omega$ $\left(G' \propto \omega^{0.38} \text{ and } G'' \propto \omega^{0.38}\right)$, while $\tan\delta$ shows no dependence on $\omega$ as shown in the figures 2(b) and 2(c). Such dependence experimentally validates equation (1). At further higher times dependence of $G'$ and $G''$ on $\omega$ continues to weaken, while $\tan\delta$ shows weak but noticeable increase with increase in $\omega$, a distinctive feature associated with a gel state. Figures 2 (a) to (c), therefore clearly demonstrate various characteristic rheological signatures of the critical gelation transition for freshly prepared aqueous suspension of Laponite.

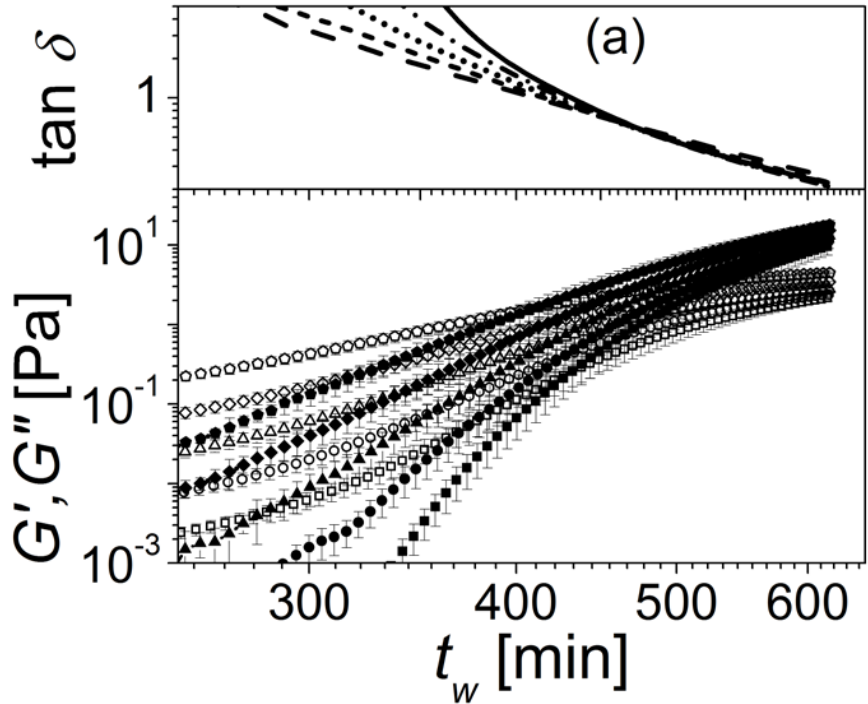

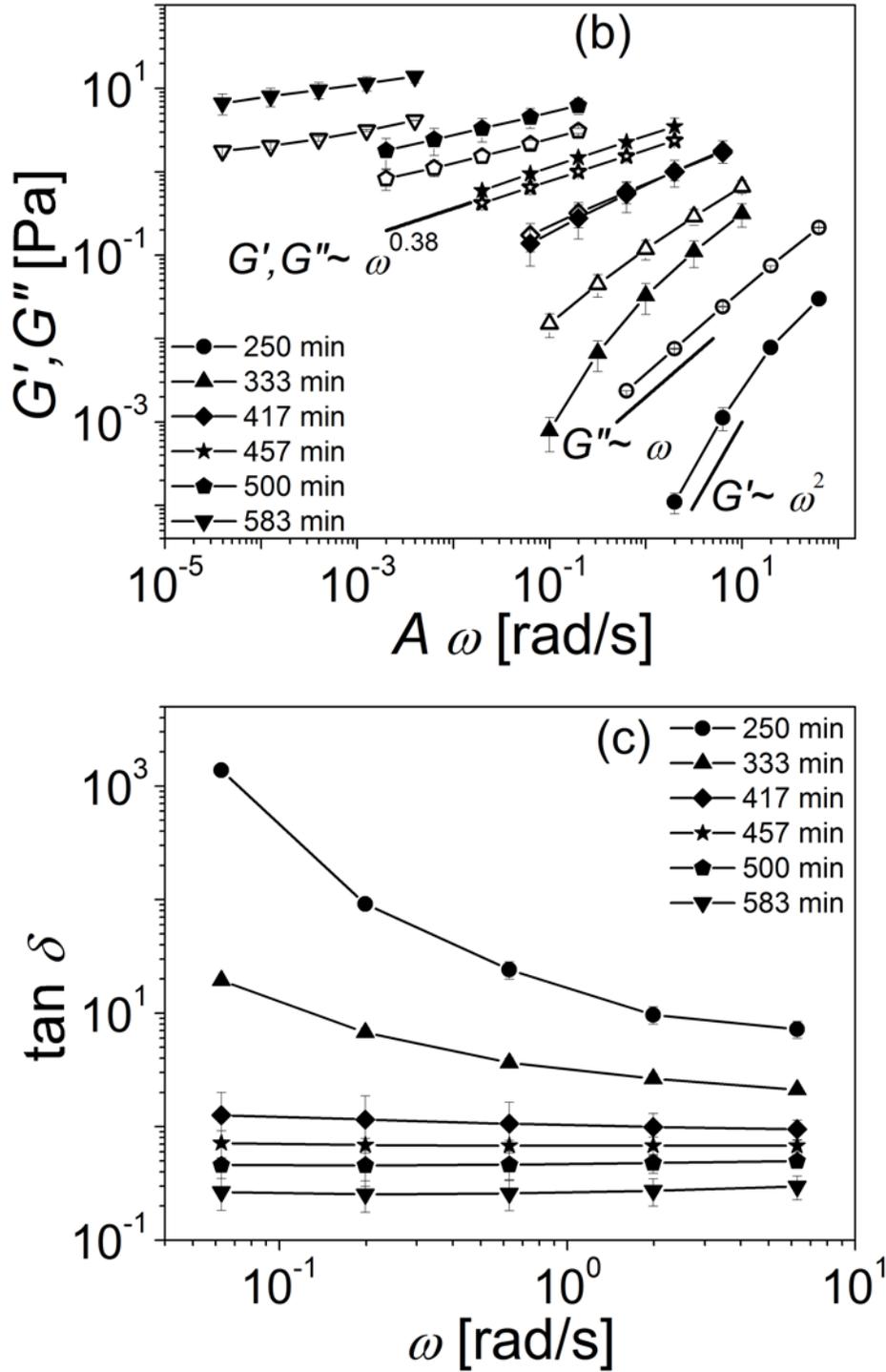

**Figure 2.** (a) Evolution of $G'$ (filled symbols), $G''$ (open symbols) and $\tan\delta$ for a freshly prepared Laponite suspension ($t_R = 0$ h) is described as a function of $t_w$ at different frequencies (for $G'$ and $G''$ from bottom to top 0.063, 0.2, 0.63, 2 and 6.3 rad/s. For $\tan\delta$, the rate of decrease weakens with increase in frequency). Owing to significant fluctuations, we have not reported behavior at 20 rad/s. In figures (b) and (c), respectively the dependence of $G'$ (filled symbols) and $G''$ (open symbols), and $\tan\delta$ are plotted as a function of frequency at different $t_w$. In figure (b) the frequency dependence at different $t_w$ is shifted horizontally for better clarity and corresponding the shift factor $A$ is mentioned in the supporting information. The lines connecting data are guide to the eye.

In figure 3(a) we plot evolution of $G'$ and $G''$ subsequent to shear melting for a sample with $t_R$ =11 h. As shown in figure 2(a), since Laponite suspension undergoes critical gelation around 7.6 h (≈457 min), 11 h old suspension is in a gel state at the time of shear melting. It can be seen in figure 1 that shear melting of a gelled suspension, although reduces its viscosity by breaking the structure but rejuvenates the same to a higher viscosity than that of associated with 0 h suspension. The subsequent evolution of $G'$ and $G''$ shows outwardly similar behavior that is observed for 0 h old suspension. However, the evolution of $\tan\delta$ at different $\omega$ is not exactly same as that observed for 0 h suspension, as various iso-frequency $\tan\delta$ evolutions do not cross at the same point, but over a narrow range of times (or values of $\tan\delta$). This behavior is more apparent in figures 3(b) and (c), where we respectively plot both the moduli ($G'$ and $G''$) and $\tan\delta$ as a function of $\omega$. It can be seen that at small $t_w$ suspension is observed to be in a liquid state. As time passes the dependence of $G'$ and $G''$ on $\omega$ weakens; however contrary to that observed for 0 h suspension, $G'$ and $G''$ do not show power law dependence on $\omega$, but cross each other. Eventually at higher $t_w$, $G'$ dominates over $G''$. Interestingly $\tan\delta$ shows continuous decrease as a function of $\omega$ suggesting a sol state over the explored range of $t_w$, except for $t_w$ =150 min, where $\tan\delta$ first decreases and then increases with $\omega$. Increase in $\tan\delta$ for certain range of $\omega$ shows presence of a gel state. We also measure evolution of $G'$, $G''$ and $\tan\delta$ after the shear melting for a sample with $t_R$ =22 h (not shown). We observe that the material shows further deviation from the classical gelation transition, wherein the range of times over which various iso-frequency $\tan\delta$ evolutions cross each other broaden compared to the behavior of sample having $t_R$ =11 h. In addition the dependence on $\tan\delta$ on $\omega$ shows continuous decrease over all the explored $t_w$, despite $G' \gg G''$ at high $t_w$.

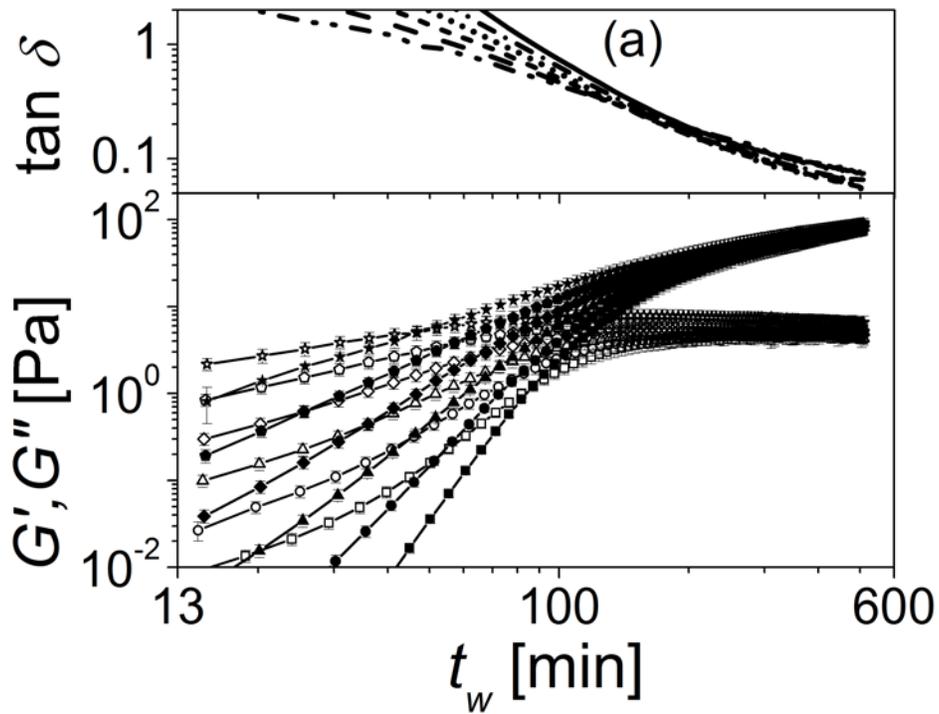

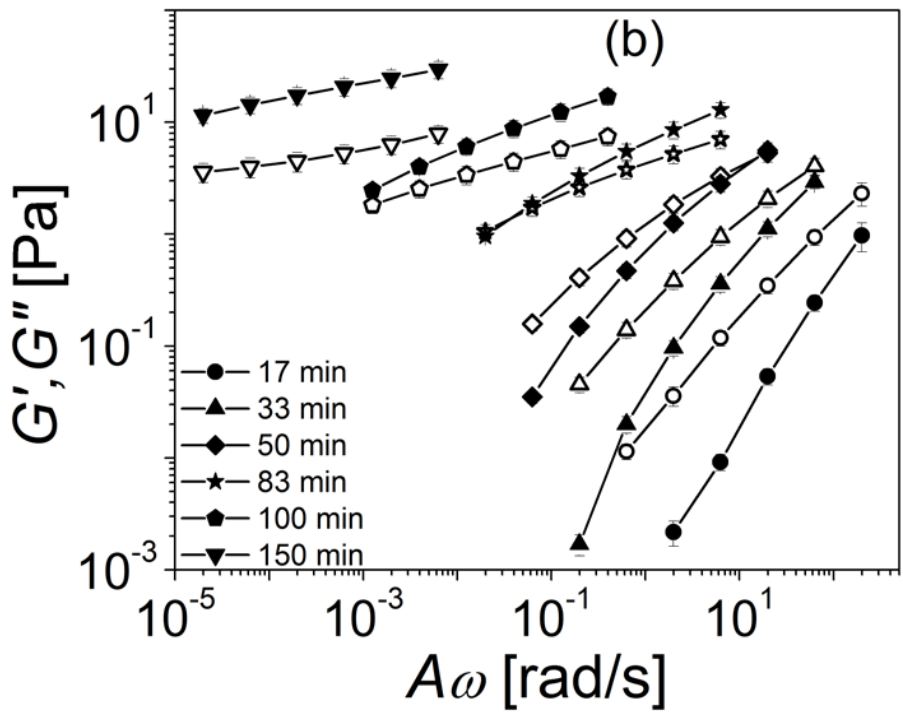

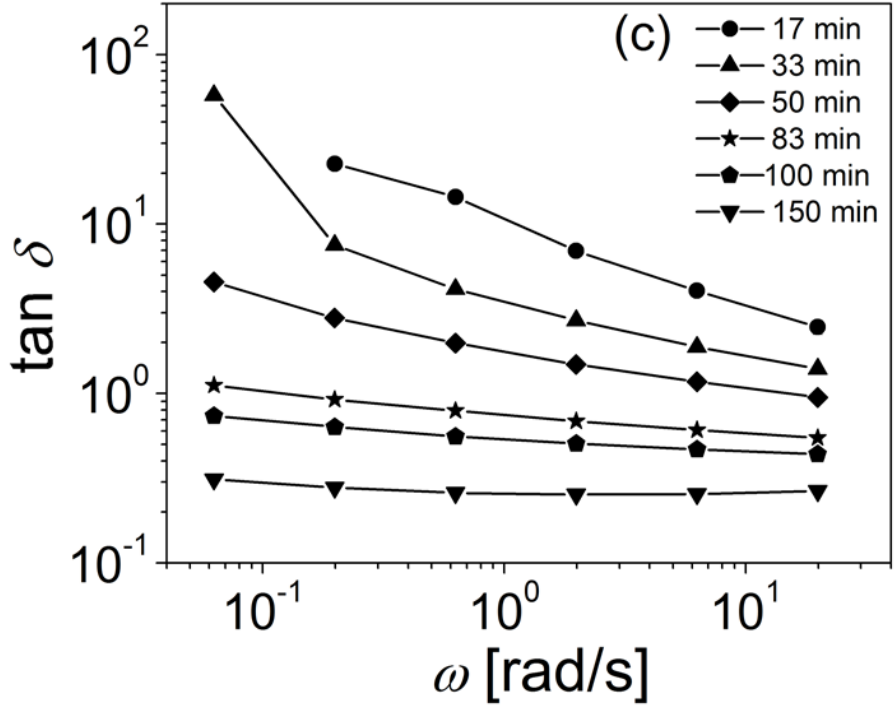

**Figure 3.** (a) Evolution of $G'$ (filled symbols), $G''$ (open symbols) and $\tan\delta$ for a Laponite suspension ($t_R = 11$ h) is described as a function of $t_w$ at different frequencies (for $G'$ and $G''$ from bottom to top 0.063, 0.2, 0.63, 2, 6.3 and 20 rad/s. For $\tan\delta$, the rate of decrease weakens with increase in frequency). In figures (b) and (c), respectively the dependence of $G'$ (filled symbols) and $G''$ (open symbols), and $\tan\delta$ are plotted as a function of frequency at different $t_w$. In figure (b) the frequency dependence at different $t_w$ is shifted horizontally for better clarity and corresponding the shift factor $A$ is mentioned in the supporting information. The lines connecting data are guide to the eye.

In figure 4 (a) we plot evolution of $G'$, $G''$ and $\tan\delta$ as a function of $t_w$ that follows shear melting for a sample with $t_R = 30$ h. It can be seen that $G'$ crosses over $G''$ at much smaller $t_w$ compared to behavior of samples with lower $t_R$. Furthermore, it can be seen that iso-frequency lines of $\tan\delta$ do not cross but approach each other in the limit of high $t_w$. In figure 4(b) we plot $G'$ and $G''$ as a function of $\omega$ (in this plot the data is plotted as it is without any shifting). It can be seen that except for two smaller times elapsed after shear melting (17 and 25 min), where $G'$ and $G''$ cross each other, $G'$ is always higher than $G''$ over the explored $\omega$. In addition, at higher $t_w$, $G''$ demonstrates a plateau as a function of $\omega$ with a hint of a minimum. As shown in figure 4(c), similar to that observed for 11 h, $\tan\delta$ decreases with $\omega$ throughout the explored range of $t_w$. At higher studied rest time ($t_R =$ 144 h), over the explored $\omega$ and $t_w$, $G'$ dominates over $G''$ as shown in figure 5(a). In addition, $\tan\delta$ is observed to decrease as a function of $t_w$ with practically identical curvature so that iso-frequency $\tan\delta$ lines remain parallel to each other. In figure 5 (b) we plot $G'$ and $G''$ as a function of $\omega$, wherein at higher times elapsed since shear melting $G'$ shows a very weak increase with $\omega$, while $G''$ shows a prominent minimum. Over this entire explored range of $\omega$ and $t_w$, $\tan\delta$ is observed to decrease with increase in $\omega$ (not shown) for $t_R =$ 144 h similar to that observed for $t_R =$ 22 and 30 h.

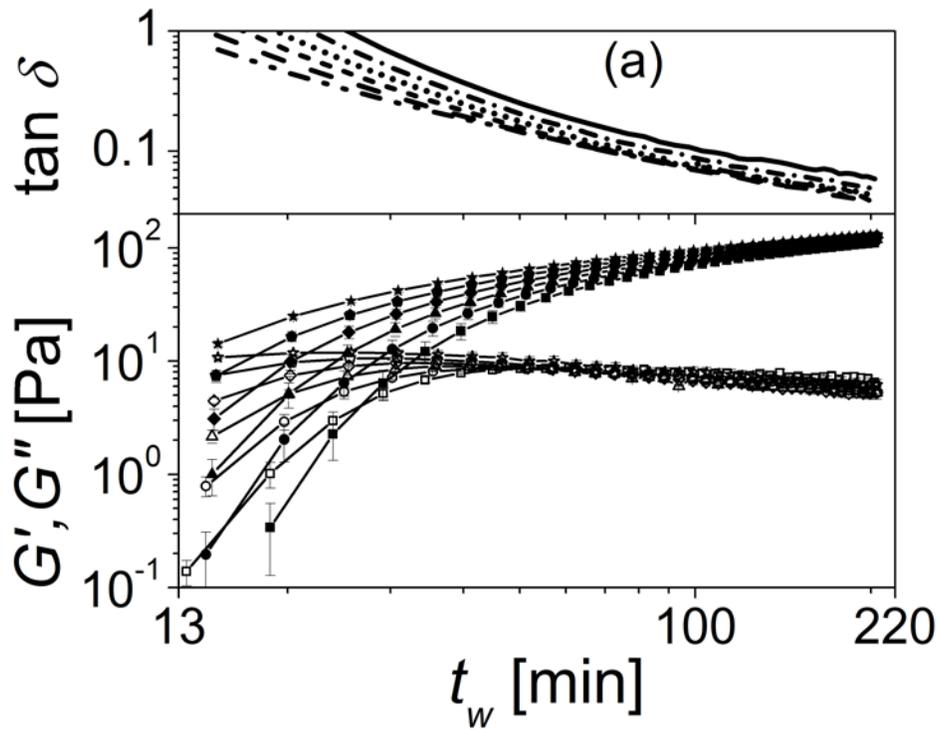

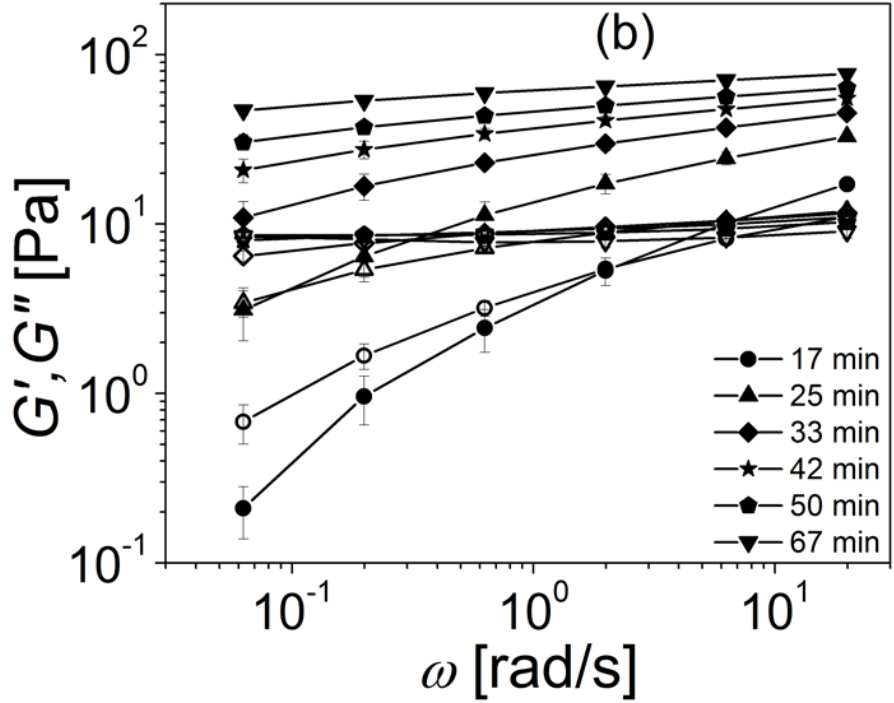

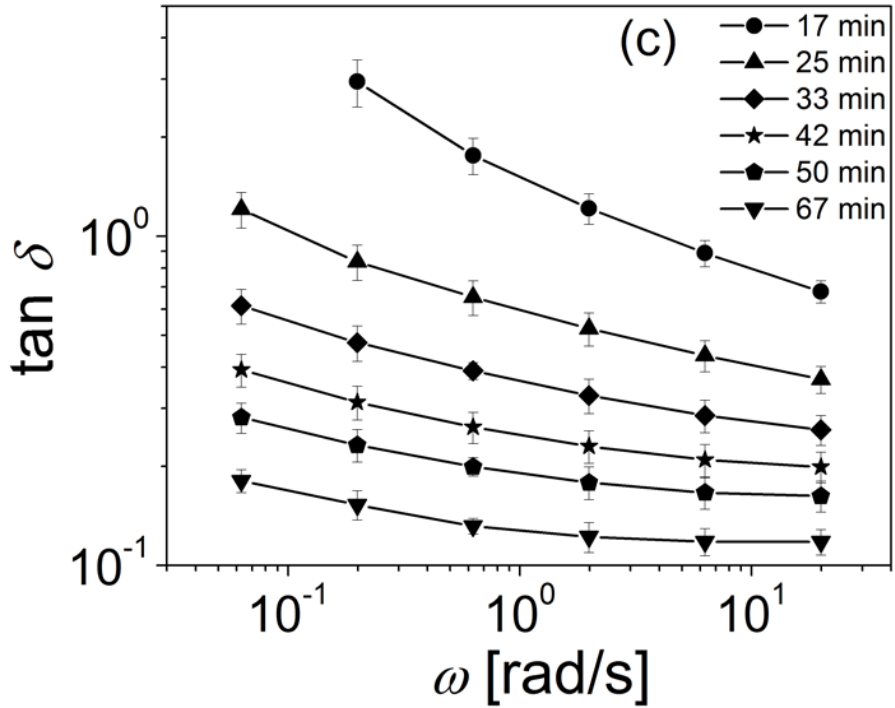

**Figure 4.** (a) Evolution of $G'$ (filled symbols), $G''$ (open symbols) and $\tan\delta$ for a Laponite suspension ($t_R = 30$ h) is described as a function of $t_w$ at different frequencies (for $G'$ and $G''$ from bottom to top 0.063, 0.2, 0.63, 2, 6.3, and 20 rad/s. For $\tan\delta$, the same types of lines are used as that of for $G'$). In figures (b) and (c), respectively the dependence of $G'$ (filled symbols) and $G''$ (open symbols), and $\tan \delta$ are plotted as a function of frequency at different $t_w$. The lines connecting data are guide to the eye.

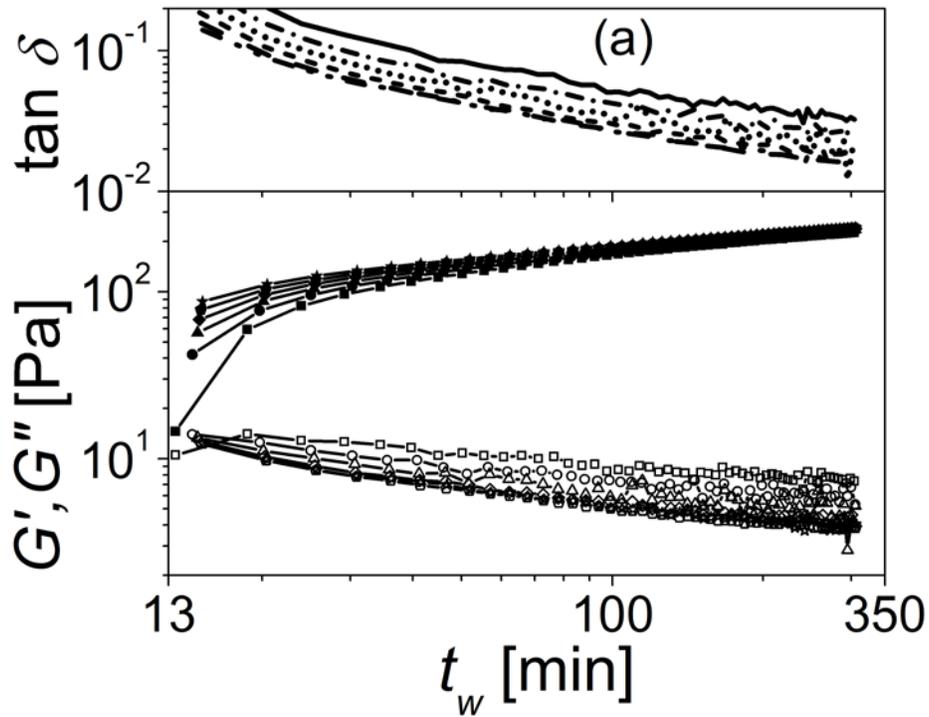

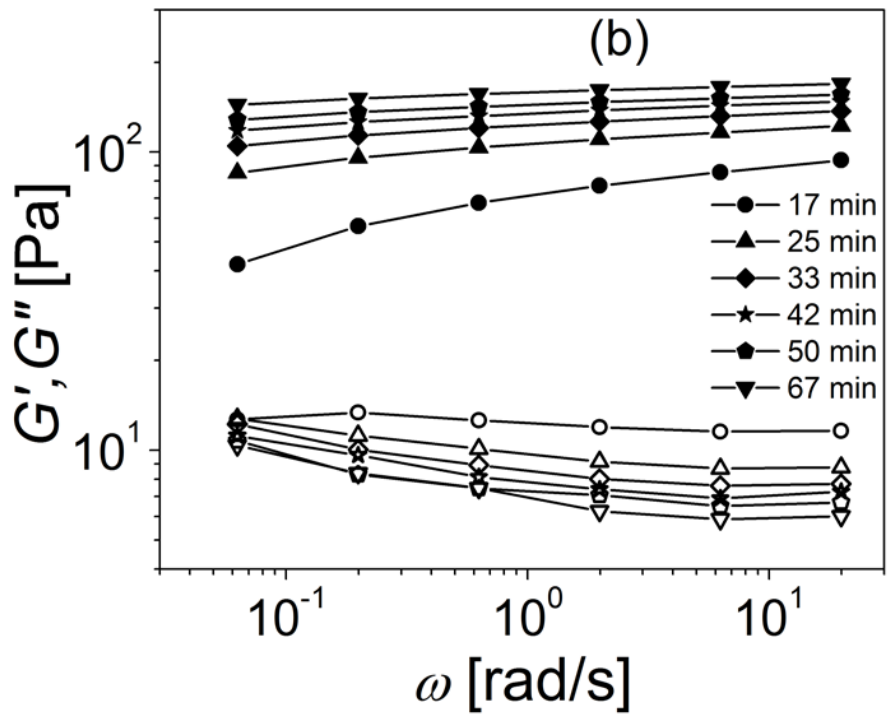

**Figure 5.** (a) Evolution of $G'$ (filled symbols), $G''$ (open symbols) and $\tan \delta$ for a Laponite suspension ($t_R = 144$ h) is described as a function of $t_w$ at different frequencies (for $G'$ and $G''$ from bottom to top 0.063, 0.2, 0.63, 2, 6.3, and 20 rad/s. For $\tan \delta$, the same types of lines are used as that of for $G'$). In figure (b) the dependence of $G'$ (filled symbols) and $G''$ (open symbols) is plotted as a function of frequency at different $t_w$. The lines connecting data are guide to the eye.

**Table 1.** Horizontal Shift factor used to shift $G'$ and $G''$ dependence on frequency.

| $t_R = 0$ h [figure 2(b)] | | $t_R = 11$ h [figure 3(b)] | |
| --- | --- | --- | --- |
| $t_w$ [min] | A | $t_w$ [min] | A |
| 250 | $10^1$ | 17 | $10^1$ |
| 333 | $10^{0.2}$ | 33 | $10^{0.5}$ |
| 417 | $10^0$ | 50 | $10^0$ |
| 460 | $10^{-0.5}$ | 83 | $10^{-0.5}$ |
| 500 | $10^{-1.5}$ | 100 | $10^{-1.7}$ |
| 583 | $10^{-3.2}$ | 150 | $10^{-3.5}$ |

From figures 2 to 5, it is clear that physical aging behavior in aqueous suspension of Laponite is strongly affected by $t_R$. In order to further investigate how rheological properties of aqueous suspension of Laponite evolve during the liquid – to – solid transition, we obtain continuous relaxation time spectra $H(\tau)$ at various times elapsed since shear melting using $G''$ dependence on $\omega$. In a continuous spectrum, $H(\tau) d \ln \tau$ represents contribution to modulus from the relaxation modes whose logarithm lie in between $\ln \tau$ and $\ln \tau + d \ln \tau$ [Ferry (1980)]. Although knowledge of $H(\tau)$ leads to exact estimation of $G'$ and $G''$, only approximate relations are available to obtain $H(\tau)$ from $G'$ and $G''$. According to Tschoegl (1989) dependence of $H(\tau)$ on $G''$ is given by:

$$H(\tau) \approx H_1''(\tau) = \frac{2G''}{\pi}\left[1 + \frac{d\ln G''}{d\ln \omega}\right]\bigg|_{\tau=1/\sqrt{3}\omega} \quad \ldots \quad \frac{d\ln H}{d\ln \tau} > 0$$

$$H(\tau) \approx H_1''(\tau) = \frac{2G''}{\pi}\left[1 - \frac{d\ln G''}{d\ln \omega}\right]\bigg|_{\tau=\sqrt{3}/\omega} \quad \ldots \quad \frac{d\ln H}{d\ln \tau} < 0$$

(3)

In equation (3), subscript 1 represents the first order approximation and involves first order derivative. The double prime associated with $H(\tau)$ simply represent that the corresponding function has been obtained from $G''$. This nomenclature is suggested by Tschoegl (1989). Incorporation equation (1) into equation (3) leads to dependence of $H(\tau)$ at GP given by: $H(\tau) \propto \tau^{-n}$. In figures 6 (a) to (d) we plot $H_1''$ as a function of $\tau$ for samples with different $t_R$ using equation (3). In figures 7, we plot $H_1''(\tau)$, at various $t_R$ but having same $t_w$ in order to analyze the effect of $t_R$ on the spectra. It should be noted that over past few decades there has been intense research to obtain more accurate form of $H(\tau)$. A brief review of the available approaches and difficulty associated with getting 'true' form of $H(\tau)$ is given by Malkin and Isayev (2006). Since obtaining $H(\tau)$ from $G'$ and $G''$ data is an inverse and moreover an ill posed problem, approach used for the determination of a spectrum is considered as a personal choice, which can lead to closer prediction of known viscoelastic functions [Malkin and Masalova (2001), Winter (1997)]. Nonetheless, more recently $H(\tau)$ is obtained by fitting a generalized (multimode) Maxwell model to $G'$ and $G''$ data [Malkin and Isayev (2006)]. However, in order to employ this approach, $G'$ and $G''$ data needs to be available over a broader range of frequencies [Winter (1997)]. In equilibrium materials the data over a broad frequency range can be easily obtained from time – temperature superposition. In the present work, however, $G'$ and $G''$ data is available only for around two decades. On the other hand, since Tschoegl's approach requires estimation of only a derivative, availability of the data over a broader frequency range is not a constraint. We also back calculated $G'$ and $G''$ using the relationship: $G^* \approx i\omega \int_{\ln \tau_{\min}}^{\ln \tau_{\max}} \tau H_1''(\tau) d\ln\tau / (1+i\omega\tau)$ [Ferry (1980)], which shows a reasonable agreement with the experimental $G'$ and $G''$ data, although the exact match is not possible. We therefore use $H(\tau)$ obtained from Tschoegl's approach for the analysis as it leads to a good representation of qualitative behavior of $H(\tau)$.

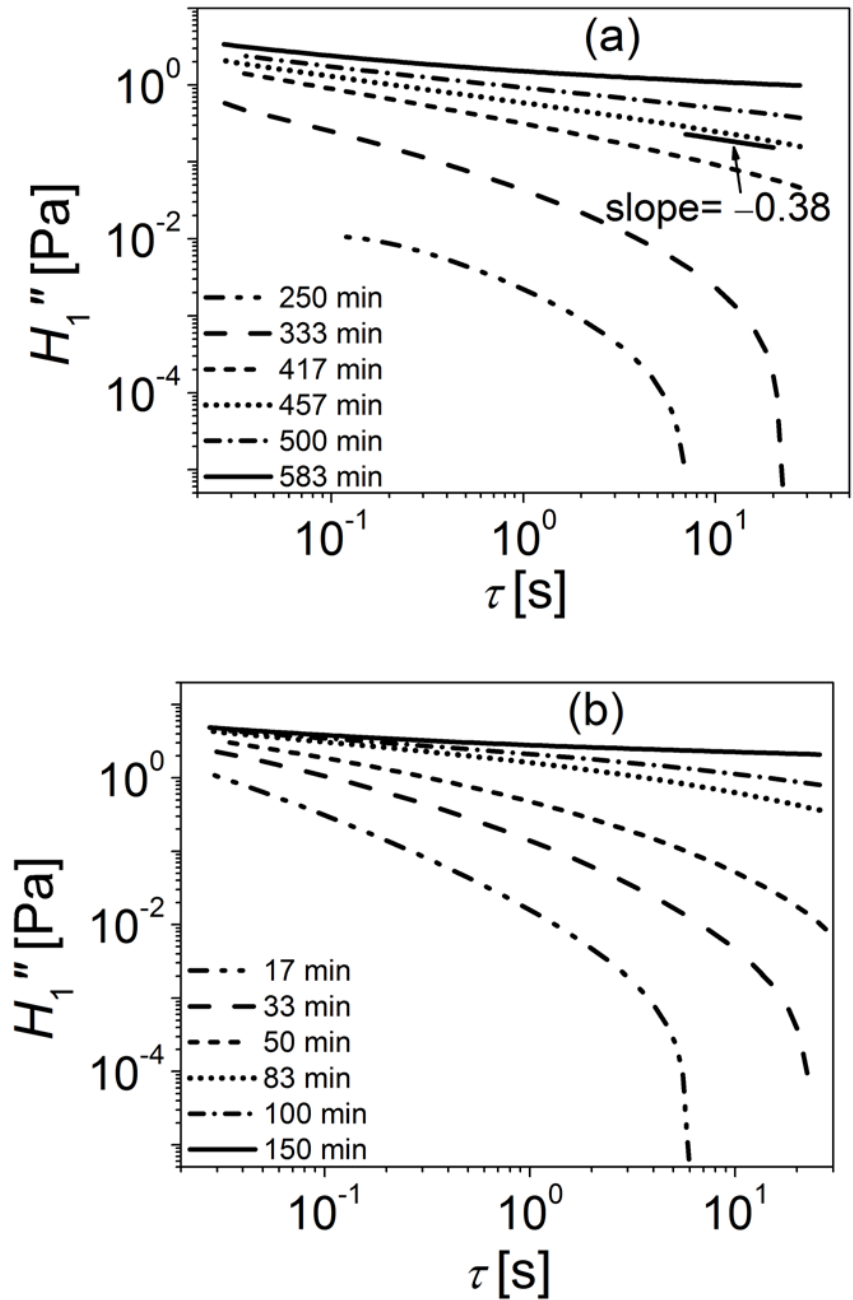

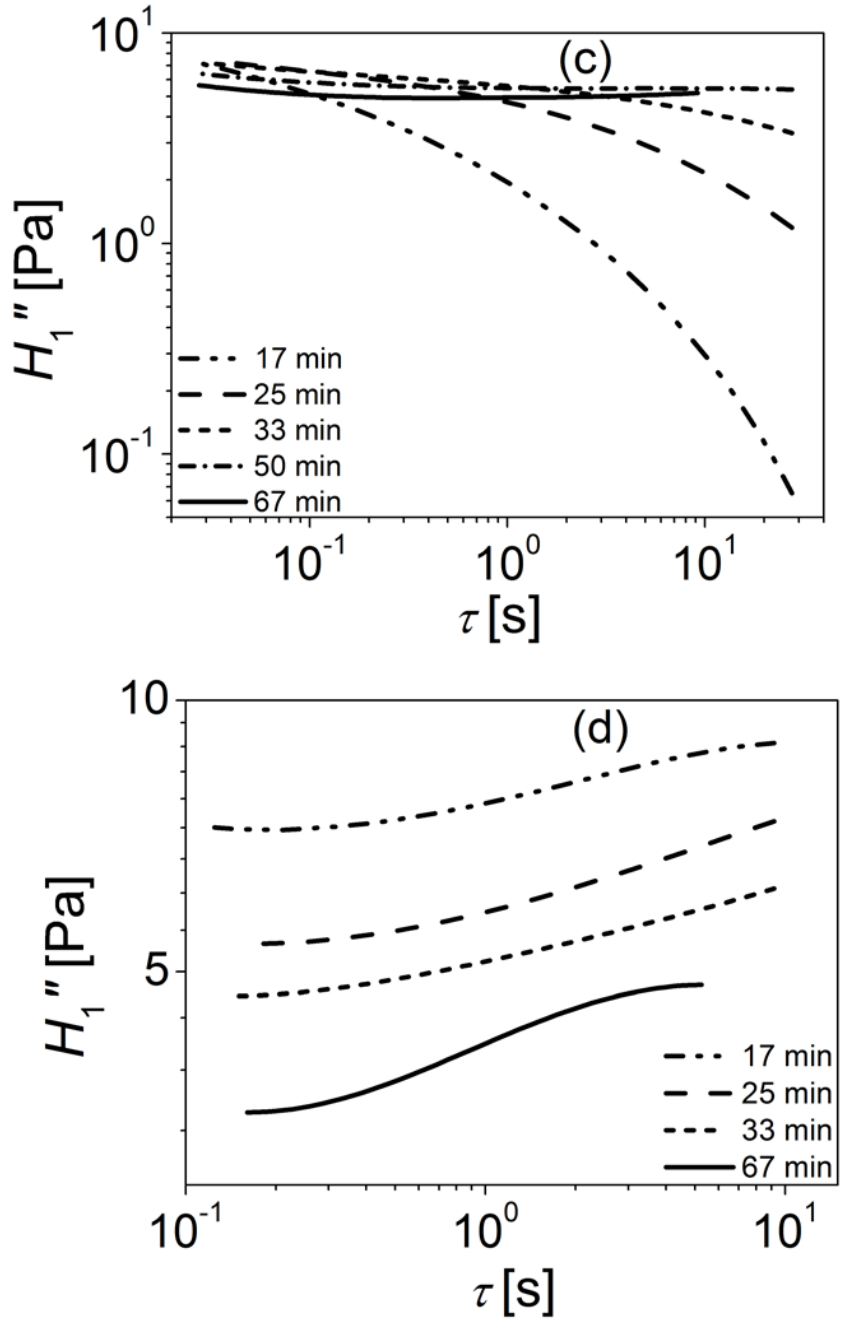

**Figure 6.** Relaxation time spectrum $H_1''(\tau)[\approx H(\tau)]$ obtained by using equation (3) at different aging time $(t_w)$ for $t_R$ =0 h (a), 11 h (b), 30 h (c) and 144 h (d).

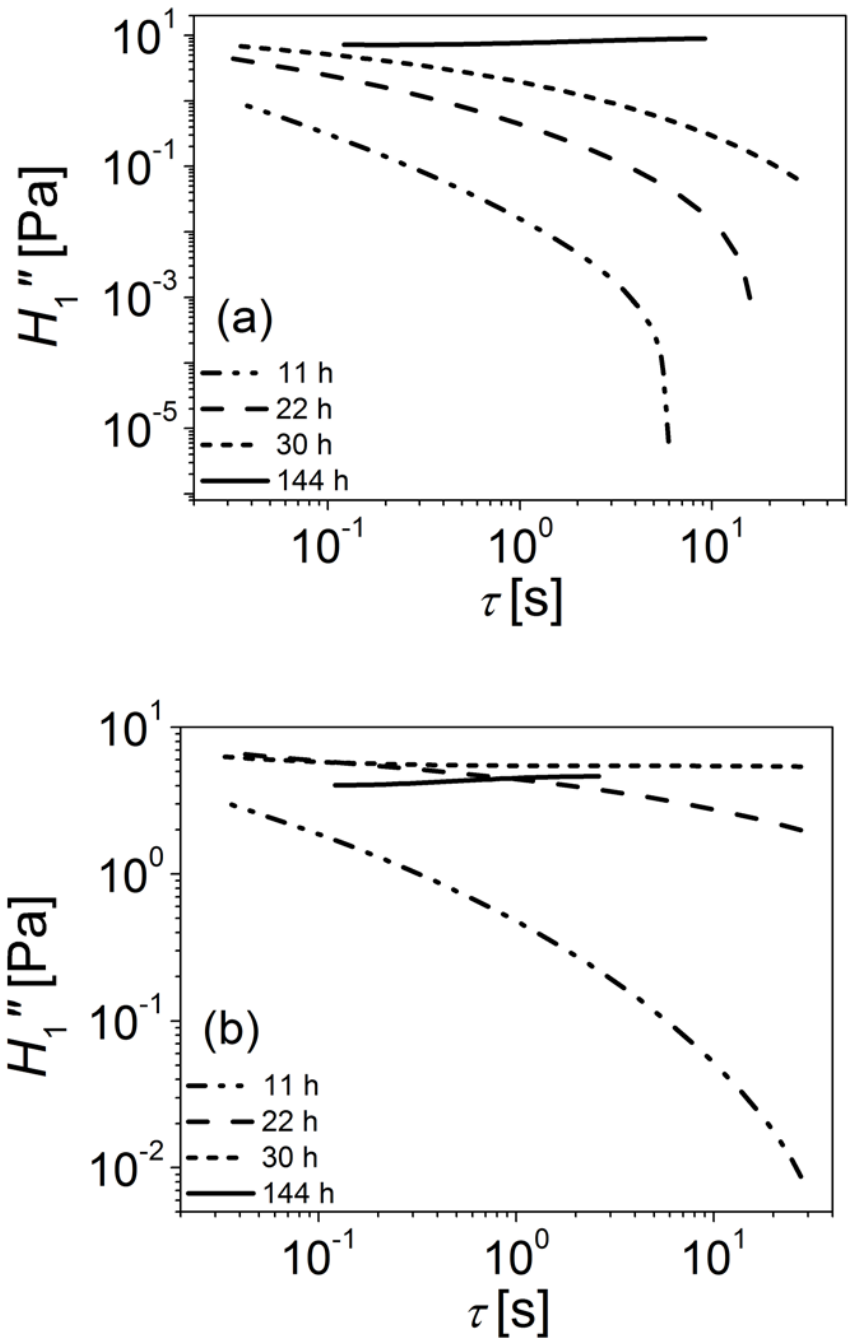

**Figure 7.** Relaxation time spectrum for $t_w$ =17 min (a), and 50 min (b) for different $t_R$.

The relaxation time spectra plotted in figures 6 and 7 provide important insight into the physical behavior responsible for variation of $G'$, $G''$ and $\tan\delta$ shown in figures 2 to 5. Below we discuss possible mechanism that could lead to the observed behavior of aqueous suspension of Laponite. In a powder form, Laponite disks are present in the form of tactoids. Soon after dispersing Laponite in water maintained at pH 10, tactoids disintegrate to smaller size or to a particle level under intense stirring [Joshi (2007)]. Furthermore, dissociation $Na^+$ ions from the faces of the particles/tactoids render the same a permanent negative charge [Joshi (2007)]. On the other hand, owing to dissociation of $OH^-$ ions, the edge acquires a positive charge [Tawari et al. (2001)]. Such charges distribution leads to edge - to - face attractive as well as face - to - face repulsive interactions among the Laponite particles [Cummins (2007)]. Such interactions lead to formation of local structures among Laponite particles/tactoids, which grow as a function of time to mesoscopic length-scales. Using rheological tools it is, however, not possible to speculate the precise nature of such structures. Nonetheless, we believe that both the attractive as well as repulsive interactions are responsible for the same [Ruzicka *et al.* (2010), Shahin and Joshi (2012)]. We hereafter term such meso-structures as aggregates. The aggregates progressively grow in size and eventually touch each other and percolate. In similar lines to that of chemical gels, the time at which the percolation occurs to span the sample space can be considered as GP. In figure 6(a) we plot $H_1''$ associated with the data shown in figure 2(a) for a freshly prepared suspension ($t_R = 0$ h). It can be seen that at small times, contribution to relaxation modulus is mainly from the fast relaxation modes that could arise from number of disconnected aggregates formed by Laponite particles. With increase in time the spectrum shifts to accommodate higher modes suggesting growth of the aggregates. However, as expected at GP, $H_1''$ indeed shows dependence on $\tau$ given by: $H_1'' \propto \tau^{-0.38}$, suggesting percolation of the aggregates. Percolated structure with relaxation spectrum having negative power law slope indicates longest mode to be the weakest as expected at GP [Winter (2013)]. After GP, $H_1''$ shows a plateau in the limit of higher relaxation modes due to enhanced population of junctions in the percolated network. Dependence of $\tan\delta$ on $\omega$ supports this scenario, where a sol, the critical gel and a gel state are clearly evident.

Once the space spanning percolated structure is formed and the material has entered a gel state, application of strong deformation field in a shear melting step induces fluidity in the material by breaking the structure. However, shear melting is unable to break the aggregates developed during the rest period down to the particle/tactoid level. Consequently, the Laponite suspension in the post shear melting scenario constitutes solid pockets of aggregates, which cannot be broken, surrounded by fluidized suspension (Joshi and coworkers, 2012). Owing to presence of unbroken aggregates (solid pockets) whose size and volume fraction increases with $t_R$, plateau viscosity of Laponite suspension increases with increase in $t_R$ as shown in figure 1. Consequently, upon cessation of the shear melting physical aging starts at a mature level wherein aggregates are already present in the suspension. This behavior is apparent from figures 6 (b) to (d) and figure 7(a), wherein contribution from the slower modes to $H_1''$ at $t_w$ =17 min can be seen to be progressively increasing with increase in $t_R$. In figure 6 (b), the spectra are plotted for $t_R$ =11 h at various $t_w$. It can be seen that with increase in $t_w$, $H_1''$ broadens to include progressively longer modes, which could be because of growth of aggregates. However unlike that observed for $t_R$ =0 h samples at $t_R$ =11 h does not show a clear GP, as various iso-frequency $\tan\delta$ lines intersect over a range of times as shown in figure 3(a). Therefore, it could be possible that, owing to presence of unbroken aggregates along with fluidized suspension, this system ($t_R$ =11 h) undergoes gelation without sampling the critical state as suggested in the literature [Krishna Reddy et al. (2012)].

For sample having $t_R$ =30 h, $H_1''$ shows a qualitatively similar behavior as demonstrated by sample having $t_R$ =11 h for low $t_w$ as shown in figure 6 (c). At high $t_w$, however, dependence of $H_1''$ on $\tau$ shows a negative slope for the fast modes (small values of $\tau$), while a positive slope for the slow modes (large values of $\tau$). This leads to a shallow minimum in the dependence of $H_1''$ on $\tau$. Such relaxation time spectra describing a minimum is very peculiar, wherein spectrum at the mentioned aging times ($t_w$ =50 and 67 min) is influenced by both short range as well as long range structures. This behavior could be attributed to contributions from fluidized paste to fast relaxation modes and from aggregates to slow relaxation modes.

Eventually at $t_R =144$ h, a complete inversion of the spectra occurs, which we plot in figure 6(d), wherein $H_1''$ is observed to increase with $\tau$ throughout the analyzed domain of relaxation timescales. Such behavior clearly indicates dominance of the long range structures which possess slow relaxation modes.

In figure 7(b) we plot effect of $t_R$ on the spectrum at $t_w=50$ min. It can be seen that for $t_R =11$ and 22 h the spectrum is dominated by fast modes. For $t_R=30$ h, spectrum is almost flat with a hint of a minimum, while for $t_R=144$ h spectrum is dominated by slow modes. In addition, for $t_R=30$ h and 144 h, $\tan\delta$ decreases with increase in $\omega$ over the entire range of explored $\omega$ and $t_w$, suggesting absence of gel state even though at high $t_w$, when $G'$ dominates over $G''$. Interestingly for a range of $t_w$ for which $H_1''$ either partly or completely inverts (or shows positive slope), $G''$ is observed to demonstrate a minimum when plotted against $\omega$ (shown in figures 4(b) and 5(b)). Presence of minima in $G''$ - $\omega$ dependence, when $G'$ shows a weak increase with $\omega$ is also observed for concentrated emulsions in a soft glassy state [Mason *et al.* (1997)]. Increase in $G''$ with decrease in $\omega$ in the low $\omega$ limit can therefore be attributed to configurational arrangements of solid pockets associated with the aggregates, while at high $\omega$ rise in $G''$ can be ascribed to viscous effect from the fluidized suspension paste surrounding the same.

The origin of the inversion of relaxation time spectra, particularly for samples kept under quiescent conditions for a longer period of time ($t_R=30$ and 144 h), is in how the structure is broken in the shear melting step. As mentioned above, we believe that with increase in $t_R$ the concentration of unbroken aggregates increases so that for the largest rest times studied here ($t_R=144$ h), the solid pockets of unbroken aggregates occupy the significant volume fraction. As a result, upon cessation of shear melting, the suspension is essentially comprised of aggregates arrested in physical cages formed by the surrounding aggregates separated by fluidized pasty suspension leading to inversion of the spectra wherein slow timescales dominate the behavior. This scenario is also reminiscent of jamming transition upon removal strong deformation field [Liu and Nagel (1998)] due to the arrested aggregates of unbroken meso-structures, rather than the particles themselves.

Interestingly Negi and coworkers' (2014) observation of evolution of $G'$, $G''$ and $\tan\delta$ for 3.5 weight % Laponite suspension, shear melted after $\left(t_R = \right)$ 4 days is qualitatively similar to that reported in figures 4 and 5. They have termed this behavior as colloidal glass as $\tan\delta$ shows decreasing dependence on frequency unlike that of in gelation transition, wherein $\tan\delta$ increases with frequency after the gel point. Interestingly Winter also proposed that increasing dependence of $H(\tau)$ on $\tau$ suggests dominance of slow timescale usually present in materials at and beyond glass transition. We believe that whether to call shear melted aged Laponite suspension, which contains significant concentration of aggregates (or solid domains) which shows increasing dependence of $H(\tau)$ on $\tau$, a glass or not is a matter of terminology. The fact still remains that the observed behavior is indeed rheological inverse of that observed for a material undergoing colloidal gelation.

This work, therefore, clearly indicates that freshly prepared 2.8 weight % aqueous suspension of Laponite having 3 mM NaCl, which is claimed to be forming a repulsive Wigner glass [Ruzicka and Zaccarelli (2011)], demonstrates all the rheological signatures of gelation transition through the critical state. This behavior is similar to that observed by 1 weight % Laponite suspension [Cocard et al. (2000)], which is accepted in the literature to undergo physical gelation through interparticle attractive interactions. Shear melted aged suspensions, however demonstrate progressive deviation from classical gelation with increase in rest time. At high rest times the material behavior is dominated by solid domains formed by unbroken aggregates present in significant concentration surrounded by fluidized pasty suspension.

### IV. Conclusion

In this work we study the aging behavior of freshly prepared and shear melted aqueous suspension of Laponite by applying oscillatory flow field having different frequencies. The samples are preserved under quiescent conditions over a rest time before applying the shear melting. We study the dependence of $G'$, $G''$ and $\tan\delta$ on frequency at different rest times, and also obtain the continuous relaxation time spectra. Overall we observe that the aging behavior of freshly prepared suspension is identical to that observed for crosslinking polymers undergoing chemical gelation. However, with increase in the rest time before shear melting, the suspensions demonstrate progressive deviation from the classical gelation. Eventually, shear melting after high rest times material shows various rheological features such as decreasing dependence of $\tan\delta$ on frequency and positive slope of relaxation time spectrum, which are opposite of that of observed for material undergoing gelation. We propose that in a freshly prepared Laponite suspension the time dependent growth of aggregates or meso-structures causes space spanning percolated network leading to formation of a physical gel. However application of high stress during shear melting on samples already in gel state does not completely destroy the structure. This creates large number of small sized aggregates that form solid domains in a continuous phase of fluidized suspension. The fraction and size of such unbroken aggregates subsequent to shear melting increase with increase in the rest time. Consequently, physical aging after the shear melting progressively deviates from the classical gelation with increase in the rest times. At large rest times, fraction of unbroken aggregates dominate such that immediately after shear melting suspension enters a solid state ($G' > G''$) demonstrating complete deviation from the gelation transition.

**Acknowledgement:** We acknowledge financial support from the department of atomic energy – science research council (DAE-SRC), Government of India.

Angelini, R., Zulian, L., Fluerasu, A., Madsen, A., Ruocco, G., and Ruzicka, B., "Dichotomic aging behaviour in a colloidal glass," Soft Matter 9, 10955-10959 (2013).

Bandyopadhyay, R., Mohan, H., and Joshi, Y. M., "Stress Relaxation in Aging Soft Colloidal Glasses," Soft Matter 6, 1462-1468 (2010).

Becker, G. W., "Mechanische Relaxationserscheinungen in nicht weichgemachten hochpolymeren Kunststoffen," Kolloid-Zeitschrift 140, 1-32 (1955).

Besseling, R., Isa, L., Ballesta, P., Petekidis, G., Cates, M. E., and Poon, W. C. K., "Shear Banding and Flow-Concentration Coupling in Colloidal Glasses," Physical Review Letters 105, 268301 (2010).

Bonn, D., Tanaka, H., Coussot, P., and Meunier, J., "Ageing, shear rejuvenation and avalanches in soft glassy materials," Journal of Physics Condensed Matter 16, S4987-S4992 (2004).

Choi, J. H., Ko, S.-W., Kim, B. C., Blackwell, J., and Lyoo, W. S., "Phase Behavior and Physical Gelation of High Molecular Weight Syndiotactic Poly(vinyl alcohol) Solution," Macromolecules 34, 2964-2972 (2001).

Cipelletti, L., and Ramos, L., "Slow dynamics in glassy soft matter," J. Phys. Cond. Mat. 17, R253–R285 (2005).

Cloitre, M., Borrega, R., and Leibler, L., "Rheological aging and rejuvenation in microgel pastes," Phys. Rev. Lett. 85, 4819-4822 (2000).

Cocard, S., Tassin, J. F., and Nicolai, T., "Dynamical mechanical properties of gelling colloidal disks," J. Rheol. 44, 585-594 (2000).

Coussot, P., "Rheophysics of pastes: A review of microscopic modelling approaches," Soft Matter 3, 528-540 (2007).

Cummins, H. Z., "Liquid, glass, gel: The phases of colloidal Laponite," J. Non-Cryst. Solids 353, 3891-3905 (2007).

Di, X., Win, K. Z., McKenna, G. B., Narita, T., Lequeux, F., Pullela, S. R., and Cheng, Z., "Signatures of Structural Recovery in Colloidal Glasses," Physical Review Letters 106, 095701 (2011).
Page | 32